\documentstyle[12pt,frascatiphys,epsfig]{article}
\begin{document}
\title{ 
OBSERVED PROPERTIES OF $\sigma$-PARTICLE
}
\author{
Muneyuki Ishida       \\
{\em Department of Physics, Tokyo Institute of Technology,
Tokyo 152-8551, Japan} \\
Shin Ishida      \\
{\em Atomic Energy Research Institite, College of Science 
and Technology}\\
{\em Nihon University, Tokyo 101-0062, Japan} \\
Taku Ishida      \\
{\em KEK, Oho, Tsukuba, Ibaraki 305-0801, Japan} \\
Kunio Takamatsu     \\
{\em Miyazaki U., Gakuen-Kibanadai, Miyazaki 889-2155, Japan} \\
Tsuneaki Tsuru     \\
{\em KEK, Oho, Tsukuba, Ibaraki 305-0801, Japan} \\
}
\maketitle
\baselineskip=14.5pt
\begin{abstract}
Recently we obtained the evidence for the 
existence of $\sigma (600)$ meson, which had been sought but 
missing for a long time,
by reanalyzing the $I=0$ $S$-wave $\pi\pi$ scattering phase shift. 
The $\sigma$-existence was also suggested
through the analyses of $\pi\pi$ production processes, 
$pp$ central collision $pp\rightarrow pp\pi^0\pi^0$ and 
$J/\psi\rightarrow\omega\pi\pi$ decay.
The observed properties through these works 
of $\sigma$ meson satisfy the mass and width
relation of $SU(2)$ linear $\sigma$ model.
The physical origin of the repulsive background phase shift $\delta_{BG}$,
which was essential to lead to the $\sigma$-existence
in our phase shift analysis,  is also due to the 
``compensating $\lambda\phi^4$-interaction" in linear $\sigma$ model.
Furthermore in this talk
the experimental property of the $\delta_{BG}$ is shown to be describable
quantitatively in the framework of linear  $\sigma$ model
including $\rho$-meson contribution.
\end{abstract}
\baselineskip=17pt
\section{Introduction}\footnote{This talk was presented by
M. Y. Ishida.}
Recently we found rather strong evidences for existence of the light 
$\sigma$-particle analyzing the experimental data obtained through
both the scattering and the production process of $I=0\ S$-wave
$\pi\pi$ system.
In the preceding talk of this conference,\cite{ref1} 
(which is referred as I in the following,) it was explained that
our applied methods of the analyses are generally consistent with the 
unitarity of $S$-matrix.
More specifically, the Interfering Amplitude(IA) method applied 
in reanalysis of $\pi\pi$ scattering phase shift satisfies the 
elastic unitarity, and  
the VMW method applied in analyses of $\pi\pi$ production processes 
is consistent with the final
state interaction theorem. 
In this talk first I collect the values of our observed property 
of $\sigma$ meson, its mass and width.

Then I show this observed property of $\sigma$ is consistent with that
to be expected in the linear $\sigma$ model. Furthermore, I shall
show, by investigating the background phase shift 
$\delta_{BG}$ theoretically in the framework of linear $\sigma$ model,
the experimental behaviors of $\delta_{BG}$ in both the $I=0$ and $I=2$
systems are quantitatively describable theoretically.
\section{Observed property of $\sigma$ meson}
{\it Reanalysis of $\pi\pi$ scattering phase shift}\ \ \ 
In the IA method\footnote{
The detailed explanation of IA method is given in I.
} the total phase shift 
$\delta_0^0$ is represented by the sum of component phase shift,
$\delta_\sigma ,\delta_{f_0(980)}$ and $\delta_{BG}$. In the actual
analysis the $\delta_{BG}$
was taken phenomenologically of hard core type: 
\begin{eqnarray}
\delta_0^0 &=& \delta_\sigma +\delta_{f_0(980)}+\delta_{BG};\ \ \ 
\delta_{BG} = -p_1r_c.
\label{eq2}
\end{eqnarray}
We analyzed 
the data of ``standard phase shift''
$\delta^0_0$ between $\pi\pi$- and K$\overline{\rm K}$-thresholds 
and also the data on upper and lower bounds
reported so far (see ref.\cite{ref2} in detail).
The results of the analyses are given in Fig. 2 of I, and
we concluded that the $\sigma$ meson exists with the property, 
\begin{eqnarray}
m_\sigma &=& 585\pm 20(535\sim 675){\rm MeV},\ \ 
\Gamma_\sigma =385\pm 70{\rm MeV}.
\label{eqexp}
\end{eqnarray}

As explained in I, the fit with $r_c=0$ corresponds to the 
conventional analyses without the repulsive $\delta_{BG}$.
In the present analysis with  $\delta_{BG}$
the greatly improved $\chi^2$ value 23.6 is obtained 
for standard $\delta_0^0$, compared with
that of the conventional analysis, 163.4. The similar
 $\chi^2$ improvement 
is also obtained for the upper and lower phase shifts.\footnote{
For upper phase shift the $\chi^2$ in the present(conventional)
analysis is $\chi^2/N_{d.o.f.}=32.3/(26-4)(135.1/(26-3))$.
For lower phase shift
$\chi^2/N_{d.o.f.}=42.1/(17-4)(111.7/(17-3))$.
} This fact strongly
suggests the existence of light $\sigma$ meson phenomenologically. 
\footnote{
Concerning this $\chi^2$ improvement Klempt gave a seemingly-strange 
criticism\cite{ref3} in his summary talk of Hadron '97, ``
However, the $\chi^2$ gain comes from a better description
of a small anomaly in the mass region around the $\rho (770)$ mass.
$\cdots$ A small feedthrough from P-wave to S-wave can very well
mimic this effect.
''
Actually the $\chi^2$ contribution from this ``anomalous'' region 
650 MeV through 810 MeV in the present(conventional) fit is
5.3(62.6), and the   $\chi^2$ contribution from the outside region is
23.6-5.3=18.3(163.4-62.6=100.8). 
Thus,  the $\chi^2$ improvement comes from a better description 
of global phase motion below 1 GeV, showing the criticism is not correct.

Furthermore, we tried to fit the data without the relevant 
data points. The obtained values of parameters
are almost equal to the ones with the data of full region, while 
obtaining the similar improvement of $\chi^2$.
}\\  
{\em Analyses of $\pi\pi$ production processes}\ \ \ 
We also analyzed the data of $\pi\pi$ production processes,
{\em pp} central collision experiment by
GAMS and 
$J/\psi\rightarrow\omega\pi\pi$ decay 
reported by DM2 collabration, 
and showed the possible evidence of the 
existence of the $\sigma$ particle. 
In the analyses we applied the 
VMW method,
where the production amplitude is represented by a 
sum of the $\sigma$, $f_0$ and $f_2$ Breit-Wigner amplitudes 
with relative phase factors. For detailed analyses, see ref. \cite{ref2}. 
The obtained mass and width of $\sigma$
are 
\begin{eqnarray}
m_\sigma &=& 580\pm 30\ {\rm MeV},\ \Gamma_\sigma =785\pm 40\ {\rm MeV}
\ \ \ {\rm for}\ pp\ {\rm central\ collision}\nonumber\\
m_\sigma &=& 480\pm 5\ {\rm MeV},\ \Gamma_\sigma =325\pm 10\ {\rm MeV}
\ \ \ {\rm for}\ J/\psi\rightarrow\omega\pi\pi\ {\rm decay}.
\end{eqnarray}
\section{Property of $\sigma$-meson and chiral symmetry}
Now the 
property of $\sigma$ meson obtained above is checked from the viewpoint 
of chiral symmetry. 
In the $SU(2)$ linear $\sigma$ model(L$\sigma$M)
the coupling constant $g_{\sigma\pi\pi}$ of the $\sigma\pi\pi$ interaction
 is related to $\lambda$ of the $\phi^4$  interaction
and $m_\sigma$ as
\begin{eqnarray}
g_{\sigma\pi\pi} &=& f_\pi\lambda =(m_\sigma^2-m_\pi^2)/(2f_\pi ).
\label{eqrel}
\end{eqnarray}
Thus, the $\Gamma_\sigma$ is related with $m_\sigma$ through the 
following equation:  
\begin{eqnarray}
\Gamma_{\sigma\pi\pi}^{\rm theor} &=& 
\frac{3g_{\sigma\pi\pi}^2}{4\pi m_\sigma^2}p_1
 \sim \frac{3m_\sigma^3}{32\pi f_\pi^2}.
\label{eqmw}
\end{eqnarray}
Substituting the experimental $m_\sigma$=535$\sim$675 MeV 
given in Eq. (\ref{eqexp})
and $f_\pi $=93 MeV into Eq.(\ref{eqmw}),
we can predict
$\Gamma_\sigma^{\rm theor}=400\sim 900\ {\rm MeV},$
which is consistent with the $\Gamma_\sigma^{\rm exp}$ given in 
Eq.(\ref{eqexp}). 
Thus the observed $\sigma$ meson may be identified with the $\sigma$
meson described in the L$\sigma$M. 

\section{Repulsive background phase shift}
{\em  Experimental phase shift and repulsive core in the I=2 system}\ \ \ 
In our phase shift analyses of the $I=0$ $\pi\pi$ system 
the $\delta_{\rm BG}$
of hard core type  introduced phenomenologically 
played an essential role.
In the $I=2$ $\pi\pi$ system, there is
no known / expected resonance, and accordingly it is expected that
the phase shift of repulsive core type will appear directly.
As shown in  Fig.~1, actually the experimental
data\cite{ref2} from the threshold to $m_{\pi\pi}\approx 1400$ MeV
of the $I=2$ $\pi\pi$-scattering $S$-wave phase shift
$\delta_0^2$ are apparently negative, and fitted well also by the
hard-core formula
$\delta_0^2=-r_c^{(2)}|{\bf p}_1|$ with the core radius of
$r_c^{(2)}=0.87\ {\rm  GeV}^{-1}$ (0.17 fm).\\
\begin{figure}[t]
 \epsfysize=4.0 cm
 \centerline{\epsffile{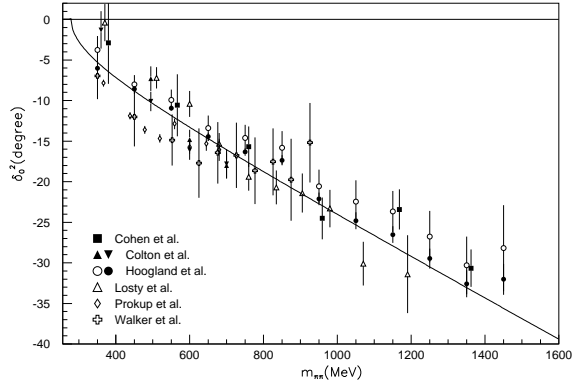}}
\caption{\it $I$=2 $\pi\pi$-scattering phase shift. Fitting by
hard core formula is also shown.}
\label{fig:i2}
\end{figure}
{\em Origin of the $\delta_{BG}$ }\ \ \ 
The 
origin of this $\delta_{\rm BG}$ seems to  have a close connection to 
 the $\lambda\phi^4$-interaction in L$\sigma$M\cite{ref2}:
It represents a contact zero-range interaction
and is strongly repulsive both in the $I=0$ and 2 systems,
and has plausible properties
as the origin of repulsive core.

The $\pi\pi$-scattering $A(s,t,u)$-amplitude  by $SU(2)$ L$\sigma$M 
is given by 
$
A(s,t,u) = (-2g_{\sigma\pi\pi})^2/(m_\sigma^2-s)-2\lambda .
$
Because of the relation (\ref{eqrel}),
the dominant part of the amplitude due to 
virtual $\sigma$ production( 1st term)
is 
cancelled by that due to 
repulsive $\lambda\phi^4$ interaction( 2nd term) 
in $O(p^0)$ level, 
and 
the $A(s,t,u)$ is rewritten into the following form:
\begin{eqnarray}
A(s,t,u) &=& \frac{1}{f_\pi^2}
\frac{(m_\sigma^2-m_\pi^2)^2}{m_\sigma^2-s}
-\frac{m_\sigma^2-m_\pi^2}{f_\pi^2}
=\frac{s-m_\pi^2}{f_\pi^2}+\frac{1}{f_\pi^2}
\frac{(m_\pi^2-s)^2}{m_\sigma^2-s},
\label{eq:Acancel}
\end{eqnarray}
where
 in the last side 
the $O(p^2)$ Tomozawa-Weinberg amplitude
and the $O(p^4)$ (and higher order) correction 
term are left.
As a result the derivative coupling property
of $\pi$-meson as a Nambu-Goldstone boson is
preserved.
In this sense the $\lambda\phi^4$-interaction can be
called a ``compensating" interaction for
$\sigma$-effect.

Thus the strong cancellation
between the positive $\delta_\sigma$  
and the negative $\delta_{\rm BG}$ in our analysis 
leading to the $\sigma$,
as shown in Fig. 2 of I, is reducible 
to the relation 
Eq.(\ref{eq:Acancel}) in L$\sigma$M.

In the following we shall make a theoretical estimate of
$\delta_{BG}$ in the framework of L$\sigma$M. 
The scattering ${\cal T}$ matrix
consists of a resonance part ${\cal T}_R$ and of a background 
part ${\cal T}_{BG}$. The  ${\cal T}_{BG}$ corresponds to the 
contact $\phi^4$ interaction and the exchange of the relevant 
resonances. The main term  of ${\cal T}_{BG}$ 
comes from the $\lambda\phi^4$
interaction. This  ${\cal T}_{BG}$ has a weak $s$-dependence
in comparison with that of  ${\cal T}_R$. 
The explicit form of   ${\cal T}_{BG}$ 
for $I=0$ and $I=2$ $S$-wave channels are given by 
\begin{eqnarray}
{\cal T}_{BG;S}^{I} &=& -6\lambda a 
+2\left( \frac{(-2g_{\sigma\pi\pi})^2}{4p_1^2}
ln\left( \frac{4p_1^2}{m_\sigma^2}+1\right)-2\lambda\right)\nonumber\\
 & + & b\cdot 2g_\rho^2\left(-1+\frac{2s-4m_\pi^2+m_\rho^2}{4p_1^2}
ln\left(\frac{4p_1^2}{m_\rho^2}+1\right)-\frac{s+2p_1^2}{m_\rho^2}
\frac{\Lambda^2+4m_\pi^2}{\Lambda^2+s}\right) ,
\label{eqB}
\end{eqnarray}
where $(a,b)=(1,2)$ for $I=0$ and  $(a,b)=(0,-1)$ for $I=2$.
Here we introduced the $\rho$ meson
contribution,which are supposed to be described by
Schwinger-Weinberg Lagrangian, \footnote{The derivative $\phi^4$
interaction appearing in Schwinger -Weinberg Lagrangian 
makes Eq.(\ref{eq9}) divergent. 
Thus we introduce a form factor with cut off 
$\Lambda\simeq 1$ GeV. }
$
{\cal L}_\rho =g_\rho\rho_\mu\cdot (\partial_\mu\phi\times\phi)
-g_\rho^2/2m_\rho^2\ (\phi\times\partial_\mu\phi )^2.
$  
In order to obtain $\delta_{BG}$ theoretically, we 
unitarize ${\cal T}$ by using the  N/D method,
${\cal T}_{BG}(s)^I=e^{i\delta_{BG}}
{\rm sin}\delta_{BG}/\rho_1=N_{BG}^I/D_{BG}^I$. We take the Born term 
Eq.(\ref{eqB}) as $N$-function. In obtaining $D$-function one 
subtraction is necessary,
\begin{eqnarray}
N_{BG}^I &=& {\cal T}_{BG;S}^I,\ \ D_{BG}^I=1+b_I
+\frac{s}{\pi}\int_{4m_\pi^2}^\infty\frac{ds'}{s'(s'-s-i\epsilon)}
\rho_1(s')N_{BG}^I(s').
\label{eq9}
\end{eqnarray}
We adopt the subtraction condition\footnote{
By using this condition the resulting $\delta_{BG}$ takes the same
value as the one obtained by simple ${\cal K}$ matrix unitarization
at the resonance energy $\sqrt{s}=m_\sigma$.}
$
Re\ D_{BG}^I(m_\sigma^2) = 1.
$
The $m_\sigma$ is fixed with the value of the best fit, 0.585 GeV;
and the values of $m_\rho$ and $g_\rho$ are 
determined from the experimental
property of $\rho$ meson; The obtained $\delta_{BG}^{I=0,2}$
is shown in Fig. 2.
\begin{figure}
 \epsfysize=5.0 cm
 \centerline{\epsffile{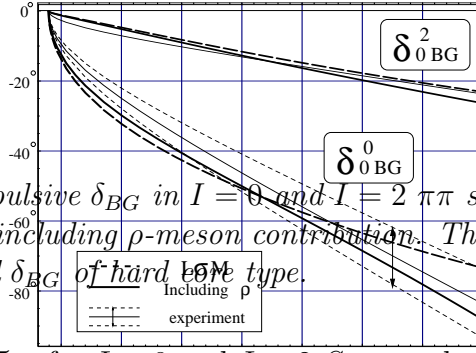}}
\caption{\it The repulsive $\delta_{BG}$ in $I=0$ and $I=2$
$\pi\pi$ scattering predicted by  the L$\sigma$M and 
by the L$\sigma$M including $\rho$-meson
contribution. The results are compared with the phenomenological 
 $\delta_{BG}$ of hard core type.}
\label{figcore}
\end{figure}
The $s$-dependence of the theoretical $\delta_{BG}$ by
L$\sigma$M including $\rho$ meson contribution is almost consistent with
the phenomenological   $\delta_{BG}$ of hard core type.
Concerning on our analysis of $\delta^{I=0}$, 
Pennington made a criticism\cite{ref4} 
that the form of $\delta_{BG}$ is completely arbitrary.
However, as shown in Fig. 2, our phenomenological
$\delta_{BG}$, Eq.(\ref{eq2}),
is almost consistent with the theoretical
prediction by L$\sigma$M. 
Thus, the criticism is not valid.
\section{Concluding remark}
We have shortly summarized the properties of the  light $\sigma$ meson
``observed'' in a series of our recent works.
The obtained values of mass and width 
of $\sigma$ satisfy the relation predicted by 
L$\sigma$M. This fact suggests the linear representation of chiral
symmetry is realized in nature.\\
In our phase shift analysis there occurrs a strong cancellation between
$\delta_\sigma$ due to the $\sigma$ resonance 
and $\delta_{BG}$, which is guaranteed by 
chiral symmetry.
A reason of overlooking $\sigma$ in conventional phase shift analysis
is due to overlooking of this cancellation mechanism.\\
The behavior of phenomenological $\delta_{BG}$ is shown to be  
quantitatively describable in the framework of L$\sigma$M including
$\rho$ meson contribution.

Finally I give a comment: By the analysis of 
$I=1/2$ $S$-wave $K\pi$ scattering phase shift in a similar method,
the existence of $\kappa (900)$ particle with a  
broad ($\sim 500$ MeV) width is suggested.
The scalars below 1 GeV, $\sigma (600)$,
$\kappa (900)$, $a_0(980)$ and $f_0(980)$ 
 are possibly to form a single scalar nonet.\cite{ref5} 
Octet members of this nonet satisfy the Gell-Mann Okubo mass formula.
Moreover, this $\sigma$ nonet is shown to satisfy the mass and width
relation of $SU(3)$ L$\sigma$M, forming with pseudoscalar $\pi$ nonet
a linear representation of chiral symmetry.
\end{document}